\documentclass[copyright,creativecommons]{eptcs}
\providecommand{\event}{PLACES 2013} % Name of the event you are submitting to
\usepackage{macron}
\usepackage{enumerate,intestazione}
\newtheorem{theorem}{Theorem}%[section]

\newtheorem{definition}{Definition}

\title{Session Types with Runtime Adaptation: \\ Overview and Examples}
\author{Cinzia Di Giusto
\institute{Universit\'e d'Evry - Val d'Essonne, Laboratoire IBISC, Evry, France}
\and
Jorge A. P\'{e}rez 
\institute{CITI and  Departamento de Inform\'{a}tica, FCT - Universidade Nova de Lisboa, Portugal}
}
\def\titlerunning{Session Types with Runtime Adaptation: Overview and Examples}
\def\authorrunning{Di Giusto \& P\'{e}rez}
\begin{document}
\maketitle

\begin{abstract}
%\emph{Adaptable processes} is a core language for concurrency 
%in which located communicating processes may be stopped, duplicated, or discarded at runtime. %along computation.
%The intention 
In recent work, we have 
developed a session types discipline 
for a calculus that features the usual constructs for session establishment and communication, 
but also two novel constructs that enable communicating processes to be stopped, duplicated, or discarded at runtime.
The aim
is to understand whether known 
 techniques for the static analysis of structured communications 
scale up to the challenging context of 
context-aware, adaptable distributed systems, in which 
disciplined interaction %following precise protocols
and 
%dynamic reconfiguration 
runtime adaptation
are intertwined concerns.
%%collections of software artifacts which communicate following precise protocols and may dynamically evolve at runtime. 
%Our work is an initial step towards this general goal; it is also 
%driven by the observation that
%%type-based 
%analyses of structured communications could be more effective %/meaningful
%if 
%%the process languages on which they are typically grounded are
%grounded on  process languages 
%enhanced to capture emerging phenomena in distributed systems.
%We focus on dynamic reconfiguration, 
%a crucial issue in, e.g., the
%%third-party, 
%cloud-based infrastructures 
%in which distributed applications are deployed nowadays.
In this short note, we summarize the main features of our session-typed framework 
with runtime adaptation, 
%in~\cite{dGP13}, 
and recall its 
basic correctness properties. We illustrate our framework by means of examples. In particular, we present a session representation of \emph{supervision trees}, a mechanism 
for enforcing fault-tolerant applications in the Erlang language.
\end{abstract}

\section{Introduction}
\label{sect:introduction}

%The analysis of structured communications using 
The type-theoretic foundations 
%has been extremely fruitful.In particular, analysis techniques based on 
provided by \emph{session types} have proved successful
in the  analysis of complex communication scenarios. 
%Numerous extensions of 
This is witnessed by, e.g., the several extensions and enhancements of 
early session types proposals~\cite{DBLP:conf/esop/HondaVK98} 
with common concepts/idioms in practical distributed programming.
%, such as, 
%%This includes, 
%e.g., forms of asynchronous~\cite{DBLP:conf/forte/KouzapasYH11} and  multiparty~\cite{DBLP:conf/popl/HondaYC08} communication. % have been integrated into session typed frameworks.
Interestingly, while often such extensions have appealed to increasingly sophisticated 
type structures (which feature, e.g., subtyping~\cite{DBLP:journals/acta/GayH05}, dependent/indexed types~\cite{DBLP:journals/corr/abs-1208-6483}, and kinding~\cite{DBLP:conf/concur/DemangeonH12}), the associated modeling language ---the polyadic $\pi$-calculus--- has remained essentially the same.
This ``asymmetric'' development 
%progress
of session types and process languages should not appear as a surprise: because of its canonicity  and expressiveness,
the $\pi$-calculus represents a rather natural %and powerful 
choice for representing 
concurrent systems which interact by following precise ---yet intricate--- behavioral patterns.
%\todo{I would add the word powerful to underline the "expressiveness" of pi: i.e. to be a rather natural and powerful choice for representing  ... }

%In recent work, we have been investigating one such patterns, namely 
We are interested in one such patterns, namely 
%\emph{dynamic reconfiguration}: 
\emph{runtime adaptation}:
it allows to represent the suspension, restarting, replacement, or abortion of running  processes. % at runtime.
As such, it can be useful to model and reason about important mechanisms in modern distributed systems, such as 
code mobility, online software update, failure recovery (as in, e.g., constructs for exceptions/compensations), %~\cite{DBLP:conf/concur/CarboneHY08,DBLP:books/sp/sensoria2011/FerreiraLRVZ11}), 
and scaling (i.e., the ability of acquiring/releasing computing resources based on runtime demand).
These mechanisms typically have a global effect over the system and are meant to be executed atomically.
Hence, they appear 
%These features make dynamic reconfiguration 
difficult to implement as $\pi$-calculus specifications:
it is not obvious how name passing/scope extrusion ---the central abstraction vehicles of the $\pi$-calculus--- can 
be used to model 
this kind of global reconfiguration primitives while retaining 
atomicity and an adequate level of abstraction for reasoning. % purposes.

Based on these limitations, %shortcomings of the $\pi$-calculus, 
and with the aim of setting a %clean, 
formal framework for reasoning about communicating systems with dynamic reconfiguration, 
with Bravetti and Zavattaro we have developed
a framework of \emph{adaptable processes}~\cite{BGPZFMOODS}. Adaptable processes extend usual %known %well-established 
process calculi %with two constructs: 
with
\emph{located processes} $\scomponent{l}{P}$ and \emph{update processes} $\adaptn{l}{U(\mathsf{X})}$, where 
$P$ is a process and 
$U$ is a %process 
context, i.e, a process with zero or more occurrences of a \emph{process variable} $\mathsf{X}$. 
%\todo{see comment below also: i would say where $U$ is a process possibly containing some occurrences of  variable $\mathsf{X}$}.
While located processes explicitly represent %a 
distributed 
%model of 
behavior, update processes 
provide a built-in adaptation mechanism
%enforce the reconfiguration of 
for located processes. %a portion of the model.
In fact, by synchronizing on name $l$,
processes $\scomponent{l}{P}$ and  $\adaptn{l}{U(\mathsf{X})}$
may reduce to  
$U\subst{P}{\mathsf{X}}$ ---the process which results from
%filling in the holes in $U$ with $P$.
substituting all free occurrences of $\mathsf{X}$ in $U$ with $P$.
%\todo{ there is a small issue here: in the sac paper and here we use variables, i am afraid it could be confusing to talk about holes, i suggest something like : $U[P]$---the process resulting by replacing variable $\mathsf{X}$ with $P$}
This way, we obtain a mechanism for dynamic process reconfiguration 
which is performed atomically in a single reduction.

In recent work~\cite{dGP13,dGP13-long}, we have investigated 
the integration of 
a session-typed discipline
into a $\pi$-calculus with located and update processes. %the two constructs above.
%dynamic reconfiguration/adaptation.
%adaptable processes. 
The intention is to understand whether 
%this well-established static analysis technique for structured communications
session types
scale up to the challenging context of 
context-aware, adaptable distributed systems, in which 
disciplined communication 
and dynamic reconfiguration are intertwined concerns.
Indeed, processes in our language
may evolve either by the usual forms of synchronization
but also by performing evolvability actions.
Hence, besides showing that well-typed processes enjoy \emph{safety} (i.e. absence of communication errors at runtime),
we have proved that typing 
entails \emph{consistency}, a property that 
%In fact, consistency 
guarantees that 
an update action for a located process is only enabled if such a process is not already involved in an active session. 
In this way, 
%Hence, by 
%guaranteeing that session-typed protocols are never disrupted by evolvability steps, 
consistency ensures a disciplined interplay of communicating and evolvability actions.
%enforced by session types.
%Technically,  %perspective, 
Our work borrows inspiration from the approach of Garralda et al~\cite{DBLP:conf/ppdp/GarraldaCD06}
on (binary) session types for Ambients, 
and extends it to the case of adaptable processes 
which run in 
arbitrary, nested distributed locations.

%More broadly, our work is %an initial step towards this general goal; it is 
%also 
%driven by the observation that
%type-based analyses of structured communications could be more effective/meaningful
%if  grounded on  process languages 
%enhanced to represent emerging phenomena in modern distributed systems.
%We have focused on dynamic reconfiguration; a similar observation may apply for other phenomena.
%%We feel this is a promising (and rather unexplored) research strand.
%%In this sense, our work may shed light on the foundations of modern programming languages.
In this presentation, we 
summarize the %main elements of the 
framework in~\cite{dGP13,dGP13-long} and illustrate it by presenting a process representation of
\emph{supervision trees}, a mechanism used in the Erlang language to enforce fault-tolerant programs~\cite{erlang:sup}.
Intuitively, supervision trees define a hierarchical structure of program components, in which 
\emph{workers} perform computation and \emph{supervisors} monitor the behavior of one or more workers.
In case of failure or unexpected termination of a worker, its supervisor should restart its behavior. 
The supervisor module in Erlang defines different \emph{restarting strategies} for supervisors.
For instance, in the \emph{one for one} restarting strategy, only the affected worker is restarted, and so supervision realizes a form of local adaptation.
In contrast, in the \emph{one for all} strategy, the supervisor restarts both the affected worker and also all of its sibling in the supervision tree, and so supervision enforces a form of global adaptation.
We show how both these strategies may be represented in our typed process framework. 
We believe the resulting models are useful to appreciate the main features of our intended model of runtime adaptation.

%------------------------------------------------------------------------------

\paragraph{Disclaimer and Organization.}
This paper is intended as a brief overview presentation to the approach and results first presented in~\cite{dGP13}.
As such, our focus here is in intuitions and examples rather than on technical details; 
the  reader is referred to~\cite{dGP13} (and to its extended version \cite{dGP13-long}) for extended details. 

The rest of the paper is structured as follows. In  Sect.~\ref{s:motex} we present a motivating example for our intended model. 
Then, Sect.~\ref{s:summa} gives a overview of our process framework for runtime adaptation with session types, and describes its main properties.
In Sect.~\ref{s:suptr} we present our model of supervision trees as session-typed processes ---both this model and the example in Sect.~\ref{s:motex} are original to this presentation.
Finally, in Sect.~\ref{s:concl} we collect some concluding remarks and discuss  directions for future work.

\section{A Motivating Example}\label{s:motex}
We begin by discussing a simple model of a \emph{workflow application}, which extends the one
%To this end, we adapt the example 
in~\cite[\S~4.2]{BGPZFMOODS}.
%where we have shown how adaptable processes are suitable in the workflow setting. 
Our model combines the main features of adaptable processes
(nested locations and update processes) 
with \emph{delegation}, the well-known mechanism for representing
reconfiguration in session-typed processes.

Briefly, a workflow is a conceptual unit that describes how  a 
number of \emph{activities} coordinate to achieve a given task.
A \emph{workflow-based application} usually consists of a \emph{workflow runtime engine} 
that contains a number of workflows running concurrently on top of it; a \emph{workflow base library} on which activities may rely on; and of a number of
 \emph{runtime services}, which are typically
 application dependent. % and implement things such as transaction handling and communication with other applications.
Exploiting nested located processes, 
in~\cite[\S~4.2]{BGPZFMOODS} we propose 
the following high-level process representation of a workflow application:
 $$
  App ~\triangleq~ \componentbig{\nm{wfa}}{\, \componentbbig{\nm{we}}{\nmu{WE} \para \nm{W}_{1} \para \cdots \para \nm{W}_{k} \para \scomponent{\nm{\nm{wbl}}}{\nmu{BL}}} \para \nmu{S}_{1} \para \cdots \para \nmu{S}_{j}\,}
% App \midef \component{\nm{wfa}}{\, \component{\nm{we}}{\nmu{WE} \parallel \nm{W}_{1} \parallel \cdots \parallel \nm{W}_{k} \parallel \component{\nm{\nm{wbl}}}{\nmu{BL}}} \parallel \nmu{S}_{1} \parallel \cdots \parallel \nmu{S}_{j}\,}
 $$

This way, 
 the application is modeled as a located process \nm{wfa} 
 which contains a workflow engine (located at \nm{we}) and a number of 
 runtime services $\nmu{S}_{1}, \ldots, \nmu{S}_{j}$.
In turn,  the workflow engine contains 
a number of workflows $\nm{W}_{1}, \ldots, \nm{W}_{k}$, a
process \nmu{WE} (which represents the engine's behavior and is left unspecified), 
and a located process \nm{wbl} representing the base library (also left unspecified).
Each workflow $\nm{W}_{i}$ (with $1 \leq i \leq k$) 
contains a process $\nmu{WL}_{i}$
(representing the workflow's logic) and is composed of $n$ \emph{activities}.
Each activity is formalized as an adaptable process $\nm{a}_{j}$ and an \emph{execution environment} $\nm{env}_j$.
The intention is that while the 
execution environment hosts the 
current (session) behavior of the activity (denoted $\nmu{P}_{j}$ below), 
location $\nm{a}_{j}$ contains an adaptation routine for location $\nm{env}_j$ (denoted $\nmu{R}_{j}$ below).
More precisely, our process representation for  $\nm{W}_{i}$  is the following:
\begin{eqnarray*}
\nm{W}_{i} & \triangleq & \componentbig{\nm{w}_{i}}{\,  \nmu{WL}_{i} \para \prod_{j=1}^n  \scomponent{\nm{env}_j}{\, \nmu{P}_{j}}  \para \scomponent{\nm{a}_j}{\,\nmu{R}_{j}\,}  \,} \quad \text{where} \\
 \nmu{R}_{j} &  \triangleq & \nopenr{a}{x}.x(b).\ifte{b = \mathtt{true}}{\big(\adaptn{\nm{env}_j}{\scomponent{\nm{env}_j}{\nmu{Q}_{j}}} \para \close{x}\big)}{\close{x}}
\end{eqnarray*}

Intuitively, $\nmu{P}_{j}$ is an unspecified session behavior that may interact with other activities, either inside or outside workflow $\nm{W}_{i}$. Process $\nmu{R}_{j}$ is meant to establish a session on name $a$ with a dual behavior contained in $\nmu{WL}_{i}$.
The purpose of such a session is very simple:  $\nmu{WL}_{i}$ will send to $\nmu{R}_{j}$ a boolean expression  which may determine 
runtime adaptation for location $\nm{env}_j$. In fact, if such an expression reduces to $\mathtt{true}$ then an update process 
 at $\nm{env}_j$ is released and the session is closed. The purpose of such an update is to replace the current behavior of $\nm{env}_j$ with process $\nmu{Q}_{j}$. Otherwise, if the boolean reduces to $\mathtt{false}$, then 
  the session between $\nmu{WL}_{i}$ and $\nmu{R}_{j}$ is closed and no modification on $\nm{env}_j$ is performed.
  
Although the possibility of adapting the behavior of session-typed processes at runtime is quite appealing, 
arbitrary update actions may jeopardize the communication protocols described by session types.
For instance, if the adaptation that $\nmu{R}_{j}$ may perform takes place while $\nmu{P}_{j}$ is in the middle of an active session, then
a communication error would arise from the resulting unmatched communication steps. Similarly, 
if $\nmu{P}_{j}$ contains replicated server definitions which $\nmu{Q}_{j}$  does not implement, then overall service availability at the workflow level would be affected.
It then becomes important to add a certain discipline to update actions, in such a way that 
(i) prescribed communication behaviors are not disrupted by them, and 
(ii) service definitions are always uniformly updated. In our session-typed process framework with runtime adaptation (summarized in the following section), we address these two natural requirements for disciplining update actions.

\section{A Framework of Disciplined,  Adaptable Processes}\label{s:summa}
\subsection{Syntax and Semantics}
Our process language corresponds to the extension of the usual polyadic $\pi$-calculus for binary session types (cf.~\cite{DBLP:conf/esop/HondaVK98,DBLP:journals/entcs/YoshidaV07}) with located and update processes.
Besides disjoint base sets for 
%\emph{names} $a, b, x, \ldots$,
\emph{names}, ranged over $a, b, x, y, \ldots$;
\emph{labels}, ranged over $n, n', \ldots$; and 
\emph{constants} (integers, booleans, names), ranged over $c, c', \ldots$, 
we consider 
\emph{locations}, ranged over $l, l', \ldots$;
\emph{integers}, ranged over $j,h,m,\ldots$; and 
\emph{process variables}, ranged over $\mathsf{X}, \mathsf{X}', \ldots$.
Then, \emph{processes}, ranged over $P, Q, R, \ldots$
and \emph{expressions}, ranged over $e, e', \ldots$
are given by the grammar in Fig.~\ref{f:syn}.
In the figure, we 
use $k, k'$ to denote names $x$ and \emph{(polarized) channels}~\cite{DBLP:journals/acta/GayH05,DBLP:journals/entcs/YoshidaV07}, ranged over $\cha^{p}, \cha_1^{p}, \ldots$ (with $p \in \{+,-\}$).
Channels are runtime entities, as formalized by our operational semantics (see below).

\begin{figure}[t]
$$
\begin{array}{lrl}
\text{Expressions  }e	&::= & \!\!\! c  ~~\sepr ~~ e_1 + e_2 \sepr e_1 - e_2  \sepr \ldots \\
\text{Processes  }P \!\!& ::= &   \!\!\! \nopenr{a}{x}.P    ~~\sepr~~ \nopena{a}{x}.P ~~\sepr~~ \repopen{a}{x}.P \\
    & \sepr &   \!\!\!   \scomponent{l}{P} ~~\sepr~~   \mathsf{X} ~~\sepr~~  \adapt{l}{P}{X}  \\
&\sepr &   \!\!\! \outC{k}{\tilde{e}}.P   	 ~~\sepr~~     \inC{k}{\tilde{e}}.P   	   	    ~~\sepr~~  \throw{k}{k'}.P     ~~\sepr~~    \catch{k}{d}.P 	 	     \\
&\sepr & 	  \!\!\!   
\branch{k}{n_1{:}P_1 \parallel \cdots \parallel n_m{:}P_m} ~~ \sepr~~ \select{k}{n};P \\
 & \sepr & 	    \!\!\! P_1 \para P_1 ~~ \sepr~~ \ifte{e}{P}{Q}~~\sepr~ \close{k}.P ~\sepr~\restr{k}{P} ~\sepr~\mathbf{0}
\end{array} 
$$
\caption{Process syntax.\label{f:syn}}
\end{figure}

Our process syntax contains an almost standard session $\pi$-calculus, with session passing (delegation)~\cite{DBLP:journals/entcs/YoshidaV07}
and replicated session acceptance constructs. 
Observe that we consider an explicit prefix for session closing, denoted $\close{k}$, 
which is useful to track the active sessions in a  located process.
Also, we consider only restriction over channels $k$.
Constructs related to runtime adaptation are given in the second  line of the grammar for processes.
We have located processes $\scomponent{l}{P}$ and update processes $\adapt{l}{P}{\mathsf{X}}$,
where 
$P(\mathsf{X})$ denotes a process with one or more occurrences of a process variable $\mathsf{X}$.
Locations are transparent: hence,  in $\scomponent{l}{P}$ process $P$ may 
evolve autonomously until it is updated by an update action on $l$. %$\updated{l}{X}{\Delta_1}{\Delta_2}{Q}$.
One objective of our type system is then to discipline such update actions, so as to avoid updating processes which contain open sessions.

%To this end, processes are  annotated: \jp{change here}
%while in , 
%in $\updated{l}{X}{\Delta_1}{\Delta_2}{P}$ both $\Delta_1$ and $\Delta_2$ denote type-based \emph{interfaces}---roughly, collections of 
%session types.
%\jp{change here} A process $\updated{l}{X}{\Delta_1}{\Delta_2}{P}$ will be able to update a located 
%process with interface (compatible with) $\Delta_1$ thus resulting in a (reconfigured) process with interface $\Delta_2$.
%These annotations are central to the 

The operational semantics for our process language is defined as a reduction relation ---Fig.~\ref{f:red} reports the required reduction rules.
The semantics generates and maintains \emph{runtime annotations} for locations: 
annotated located processes are written 
$\component{l}{h}{\Delta}{P}$, where  integer $h$ denotes the number of \emph{active sessions} in $P$.
These annotations are key in avoiding update actions over located processes which are currently engaged in active sessions.
Reduction rules rely on 
(syntactic) contexts (ranged over $C, D, E$), which represent the nested structure of located processes.
Such contexts are defined by the following syntax:
$$
C, D, E, \ldots ::= \{\bullet\} \sepr \component{l}{h}{\Delta}{C \para  P} %\sepr  \sepr \restr{\kappa}{C}  
$$

\begin{figure}[ht]
%$$
%\mbox
$$
\begin{array}{ll}
 \rulename{r:Open} & 
E\big\{C\{\nopena{a}{x}.P\} \para  D\{\nopenr{a}{y}.Q\}\big\} \pired  \\
& 
\hfill E^{++}_{} \big\{\restr{\cha}{\big({C^{+}_{}\{P\sub{\cha^+}{x}\}  \para  D^{+}_{}\{Q\sub{\cha^-}{y}\} }\big)\big\} } \vspace{2.0mm} \\
\rulename{r:ROpen} & 
E\big\{C\{\repopen{a}{x}.P\}  \para  D\{\nopenr{a}{y}.Q\} \big\}  \pired  \\
& 
\hfill E^{++}_{}\big\{\restr{\cha}{\big({C^{+}_{}\{P\sub{\cha^+}{x}  \para \repopen{a}{x}.P \}  \para  D^{+}_{}\{Q\sub{\cha^-}{y}\} }\big)}\big\}  
 \vspace{2.0mm}\\
\rulename{r:Upd} & 
E\big\{C\{\component{l}{0}{}{Q}\} \para  D\{\adapt{l}{P}{X}\}\big\} 
\pired  
E_{}\big\{C\{P\sub{Q}{\mathsf{X}}\}  \para  D\{\nil\}\big\} \vspace{ 1.5mm} \\
% \srulename{r:Move} & 
% E\big\{C\{\component{l}{h}{}{Q}\} \para  D\{\migrate{l}{m}\}\big\}~~ \pired   
% E_{}\big\{C\{\component{m}{h}{}{Q}\}  \para  D\{\nil\}\big\} \vspace{ 1.5mm} \\
% \srulename{r:Kill} & 
% E\big\{C\{\component{l}{0}{}{Q}\} \para  D\{\destroy{l}\}\big\}~~ 
% \pired   
% E_{}\big\{C\{\nil\}  \para  D\{\nil\}\big\} \vspace{ 1.5mm} \\
\rulename{r:I/O} &
E\big\{C\{\outC{\cha^{\,p}}{\tilde{e}}.P\} \para  D\{\inC{\cha^{\,\overline{p}}}{\tilde{x}}.Q\}\big\} 
\pired E\big\{C\{P\} \para  D\{Q\sub{\tilde{c}\,}{\tilde{x}}\}\big\} \quad (\tilde{e} \downarrow \tilde{c}) \vspace{ 1.5mm}
\\
\rulename{r:Pass} &
E\big\{C\{\throw{\cha^{\,p}}{\cha'^{\,q}}.P\} \para  D\{\catch{\cha^{\,\overline{p}}}{x}.Q\}\big\}\pired  E\big\{C^{-}\{P\} \para  D^{+}\{Q\sub{\cha'^{\,q}}{x}\}\big\} \vspace{ 1.5mm}
\\
\rulename{r:Sel} &
E\big\{C\{\branch{\cha^{\,p}}{n_1{:}P_1 \parallel \cdots \parallel n_m{:}P_m}\} \para  D\{\select{\cha^{\,\overline{p}}}{n_j};Q\}\big\} 
\pired \\
& 
\hfill E\big\{C\{P_j\}\para  D\{Q\}\big\}  \quad (1 \leq j \leq m)  \vspace{ 1.5mm}
\\
\rulename{r:Close} &
E\big\{C\{\close{\cha^{\,p}}.P\} \para  D\{\close{\cha^{\,\overline{p}}}.Q\}\big\} \pired  E^{--}\big\{C^{-}\{P\} \para  D^{-}\{Q\}\big\} \vspace{ 1.5mm}
\\
\rulename{r:IfTr} &
C\{\ifte{e}{P}{Q}\} \pired C\{P\}  \quad (e \downarrow \mathtt{true})  \vspace{ 1.5mm}
\\
 \rulename{r:IfFa} &
C\{\ifte{e}{P}{Q}\} \pired C\{Q\}  \quad (e \downarrow \mathtt{false}) \vspace{ 1.5mm}
\\

 \rulename{r:Str} &
\text{if } P \equiv P',\, P' \pired Q', \,\text{and}\, Q' \equiv Q ~\text{then} ~ P \pired Q \vspace{ 1.5mm}
\\
 \rulename{r:Par} & \text{if } P \pired P' ~~\text{then} ~~ P \para Q \pired P' \para Q  \vspace{ 1.5mm}
\\
 \rulename{r:Res} &
\text{if } P \pired P' ~~\text{then} ~~ \restr{\cha}{P} \pired \restr{\cha}{P'} %\vspace{2mm} \\

\end{array}
$$

\caption{ Reduction Rules. We use $\tilde{e} \downarrow \tilde{c}$ to denote an evaluation relation on expressions.\label{f:red}}
\end{figure}

This way, given a context $C\{\bullet\}$ and a process $P$, 
we write $C\{P\}$ to denote the process obtained by filling in the holes in $C$ with $P$.
The intention is that $P$ may reduce inside $C$, thus reflecting the transparent containment realized by location nesting.
The semantics relies on
a structural congruence relation (omitted), which handles scope extrusion 
for channels  in the usual way. 
Rules 
 \rulename{r:Open} and \rulename{r:Close}, which formalize session establishment and termination, resp., use two operations over contexts, denoted 
$C^{+}$ and $C^{-}$. Informally, %given a context $C$, 
$C^{+}$ (resp. $C^{-}$) denotes the context obtained from $C$ by 
 increasing (resp. decreasing)
the annotation $h$ in all of its located processes. %\jp{mention runtime channels and endpoints}
%Hence, $C^{+}$ (resp. $C^{-}$) propagates the information that a new session has been opened (resp. closed).
%both operations are defined inductively on the structure of nested, located processes.
Rule \rulename{r:Upd} captures the essence of our notion of consistency:
a (located) process $Q$ can be updated only if it contains no active sessions.
The other rules are completely standard, generalized to our setting of nested, transparent locations.

%and (ii)~its interface (denoted $\intf{Q}$ in the rule) is \emph{compatible} with the update process. 
%This is denoted $\intf{Q} \compat\Delta_1$. 
%For the sake of generality, this compatibility relation %over interfaces 
%is left unspecified.

\subsection{Type System}
%Our type syntax is given below.
%Assuming a set of \emph{basic types}, ranged over $\capab, \capab', \ldots$, we have:
%$$
%\begin{array}{lrl}
%%\text{Basic types  } \capab & ::= & \mathsf{int} \sepr \mathsf{b ool} \sepr \dots ~~~ \\
%\text{Pre-session types  }\zeta & ::= & \epsilon  \!
%		~~\sepr~~   !(\tilde{\capab}).\zeta 
%		~~\sepr~~ ?(\tilde{\capab}).\zeta  
%		~~\sepr~~   !(\sigma).\zeta 
%		~~\sepr~~ ?(\sigma).\zeta 
%		~~\sepr~~   t 
%		~~\sepr~~   \mu t.\zeta \\
%	& \sepr & 	  \&\{n_1:\zeta_1, \dots,  n_k:\zeta_k \}   
%		~~\sepr~~   \oplus\{n_1:\zeta_1, \dots , n_k:\zeta_k \}   \\
%\text{Closed pre-session types  } \rho & \triangleq &  \{ \zeta ~|~ \mathsf{ftv}(\zeta) = \emptyset\} \\
%%\text{Type Qualifiers  } \qua & ::= & \qual \sepr \quau  \! \\
%\text{Session types  }\omega &::=& \rho_\qua \sepr 	\bot 
%\end{array}
%$$
    
\begin{figure}[ht]    
$$
\textsc{Types}
$$    
    $$
\begin{array}{lclr}
\capab, \sigma & ::= & \mathsf{int} \sepr \mathsf{bool} \sepr \dots ~~~& \text{basic types}\\

\alpha, \beta &::=  &  !(\tilde{\capab}).\alpha \sepr ?(\tilde{\capab}).\alpha & \text{send, receive} \\ 
		& \sepr &  !(\beta).\alpha \sepr ?(\beta).\alpha & \text{throw, catch} \\ 
		& \sepr &  \&\{n_1:\alpha_1, \dots,  n_m:\ST_m \}  \quad &  \text{branch} \\
		& \sepr &  \oplus\{n_1:\alpha_1, \dots , n_m:\ST_m \}  &  \text{select} \vspace{0.5mm} \\
		& \sepr &  \epsilon \!\!\!\!\!\!& \text{closed session}
\end{array}
$$

$$
\textsc{Environments}
$$
$$
\begin{array}{lclr}
		\qua & ::= &  \qual \sepr \quau  & \text{type qualifiers} \vspace{0.5mm} \\
\INT &::= &  \emptyset \sepr \INT, \langle a: \ST, h \rangle & \text{interface} \\
\ActS & ::= &  \emptyset \sepr \ActS , k:\ST \sepr \ActS ,  [k:\ST]  & \text{typing with active sessions}\\
\Gamma &::= &  \emptyset \sepr \Gamma, e:\capab \sepr \Gamma, a:\langle \ST_\qua , \overline{\ST_\qua} \rangle \quad& \text{first-order environment}\\
\Theta &::= &  \emptyset \sepr  \Theta,\mathsf{X}:\INT \sepr \Theta,l:\INT     \quad& \text{higher-order environment}
\end{array}
$$  
\caption{ Type Syntax and Typing Environments \label{f:type}}
\end{figure}

We have deliberately aimed at retaining a standard session type structure; see Fig.~\ref{f:type}. % close to standard session types presentations:
We assume a notion of \emph{duality} over session types $\alpha$, noted $\overline{\alpha}$, 
defined as usual.
We now comment on our notion of \emph{typing judgment}. 
We extend usual judgments with an explicit \emph{interface}, denoted $\INT$. 
Intuitively, interfaces 
are assignments from names to session types which 
describe the services declared in a given process.
We also consider typings $\ActS$ and environments $\Gamma$ and $\Theta$.
As typical in session types disciplines, 
typing $\ActS$ collects 
assignments from channels from session types, therefore describing 
currently active sessions. 
In $\ActS$ we also include \emph{bracketed assignments}, denoted $[\kappa^p:\ST]$,
which represent active but restricted sessions. 
As we discuss in~\cite{dGP13-long}, bracketed assignments arise in the typing of channel restriction, 
and are key to keep a precise count of the active sessions in a given location.
$\Gamma$ is a first-order environment which maps expressions to basic types
and names to pairs of \emph{qualified} session types.
Within the interface, session types can be linear ($\qual$) or unrestricted ($\quau$). 
While $\qual$ is used for those session types to be used just once, 
$\quau$ annotates those session types intended to feature a persistent behavior.
The higher-order environment $\Theta$ 
collects assignments of process variables and locations to interfaces. 
While the former kind of assignments are relevant to update processes, the latter concern located processes.
Given these environments, a \emph{type judgment} is  of form
$$\judgebis{\env{\Gamma}{\Theta}}{P}{\type{\ActS}{\INT}} $$ 
meaning that, under environments $\Gamma$ and $\Theta$, 
process $P$ has active sessions declared in $\ActS$ and interface $\INT$. 

%This way, a \emph{type judgment} is  of form
%$$\judgebis{\env{\Gamma}{\Theta}}{P}{\type{\Phi}{\Delta}} $$
%stating that, under environments $\Gamma$ and $\Theta$, 
%process $P$ has active sessions declared in $\Phi$ and interface %(a collection of not yet open sessions) 
%$\Delta$. 

We only present typing rules for adaptable and update processes; see~\cite{dGP13-long} for a full account.
%$$
%\begin{array}{lcclc}
%\rulename{t:Loc} & \cfrac{\judgebis{\env{\Gamma}{\Theta}}{P}{\type{\Phi}{\Delta}} \qquad h = \#\{c \mid c:\omega \in \Phi\} }{\judgebis{\env{\Gamma}{\Theta}}{\component{l}{h}{\Delta}{P} }{ \type{\Phi}{\Delta}}} &    &
%\rulename{t:Upd}   & 
%\cfrac{\judgebis{\env{\Gamma}{\Theta,\mathsf{X}:{\Delta_1}}}{P}{\type{\emptyset}{ \Delta_2 }}}{\judgebis{\env{\Gamma}{\Theta}}{\updated{l}{X}{\Delta_1}{\Delta_2}{P}}{\type{\emptyset}{ \emptyset}}}
%\end{array}
%$$
\begin{eqnarray*}
& \inferrule[\rulename{t:Loc} ]{
       \Theta \vdash l:\INT \quad \judgebis{\env{\Gamma}{\Theta}}{P}{\type{\ActS}{\INT' } } \quad  h = | \ActS |  \quad \INT' \intpr \INT
 }{\judgebis{\env{\Gamma}{\Theta}}{\component{l}{h}{\INT}{P} }{ \type{\ActS}{\INT' }}} 
\qquad
 \inferrule[\rulename{t:Adapt} ]{\Theta \vdash l:\INT  \quad  \judgebis{\env{\Gamma}{\Theta,\mathsf{X}:{\INT}}}{P}{\type{\emptyset}{ \INT' }}}{\judgebis{\env{\Gamma}{\Theta}}{\adapt{l}{P}{X}}{\type{\emptyset}{ \emptyset}}} & 
\end{eqnarray*}

Rule \rulename{t:Loc} types located processes and performs two checks. First, the runtime annotation 
$h$ is computed by counting the assignments (standard and bracketed) declared in $\ActS$.
Second, the rule checks that the interface of the located process is less  or equal  than the
declared interface of the given location. 
In the rule, $\intpr$ denotes a preorder over interfaces, defined in~\cite{dGP13-long}; hence, the premise in the rule informally 
ensures that the process behavior does not ``exceed'' the expected behavior within the location.
It is worth observing how a typed located processes has the exact same typing and interface of its contained process:
this is how transparency of locations arises in typing.
Rule \rulename{t:Adapt} types update processes. Observe how the interface associated to the process variable of the given 
runtime adaptation routine should match with the declared interfaces for the given location. 

\subsection{Safety and Consistency}
We now summarize the main properties of well-typed processes.
They are thoroughly developed in~\cite{dGP13-long}.
We say a typing $\ActS$ is balanced iff 
for all $\kappa^p:\ST \in \ActS$ (resp. $[\kappa^p:\ST] \in \ActS$)
then also 
$\kappa^{\overline{p}}: \overline{\ST} \in \ActS$ (resp. $[\kappa^{\overline{p}}: \overline{\ST}] \in \ActS$).
Then we have the following \emph{subject reduction} theorem:

\begin{theorem}[Subject Reduction]\label{th:subred}
If $\judgebis{\env{\Gamma}{\Theta}}{P}{\type{\ActS}{\INT}}$ with $\ActS$ balanced and $P \pired Q$ then 
 $\judgebis{\env{\Gamma}{\Theta}}{Q}{ \type{\ActS'}{\INT'}}$, for some $\INT'$ and balanced $\ActS'$.
\end{theorem}

In order to state  runtime safety, we 
require some auxiliary definitions. Process $P$ is said to be 
a \emph{$\kappa$-process}
if it sends (resp. receives) a value/session on $\kappa^p$, if it makes a selection (resp. offers a choice), or if 
it closes a session on $\kappa^p$. Then, 
two $\kappa$-processes constitute a  \emph{$\kappa$-redex} if they are complementary to each other, possibly enclosed in suitable contexts:
i.e., if one sends, the other receives on $\kappa^p$; if one makes a selection, the other chooses; or if they are closing the same session $\kappa^p$.
Then, $P$ is an \emph{error} if, up to structural congruence, it contains 
either
exactly 
two $\kappa$-processes that do not form a $\kappa$-redex
or
three or more $\kappa$-processes.
Our first result is that well-typed process do not lead to communication errors:

\begin{theorem}[Typing Ensures Runtime Safety]\label{t:safety}
If $\judgebis{\env{\Gamma}{\Theta}}{P}{\type{\ActS}{\INT}}$ with $\ActS$ balanced
then $P$ never reduces into an error.
\end{theorem}

We now state and proof \emph{consistency}, the property that ensures that update actions do not disrupt session behavior.
We require an auxiliary notation. Let $\pired_{\text{upd}}$ stand for an update action, i.e., 
a reduction inferred using rule $\rulename{r:Upd}$, possibly 
with the interplay of (structural) congruence rules.
We have:

\begin{definition}[Consistency]\label{d:consis}
A process $P$ is \emph{update-consistent} 
if and only if,
 for all $P'$ and $\kappa$ 
such that $P \pired^{*} P'$ and $P'$ contains a $\kappa$-redex, 
if $P' \pired_{\text{upd}} P''$
then $P''$ contains a $\kappa$-redex.
\end{definition}

Our type system ensures that both annotations enabling update actions  and
interface assignments to locations 
are correctly assigned and maintained along reduction. Indeed, as our second result we have:

\begin{theorem}[Typing Ensures Update Consistency]\label{t:consist}
If $\judgebis{\env{\Gamma}{\Theta}}{P}{\type{\ActS}{\INT}}$, with $\ActS$ balanced,
then $P$ is update consistent.
\end{theorem}

\section{A Session Model of Supervision Trees}\label{s:suptr}

In this section, we present encodings of %the concept of 
\emph{supervision trees} in our session language 
with located and update processes. 
As supervision trees (as provided by the supervisor module of Erlang~\cite{erlang:sup}) 
define a basic principle for designing and programming fault-tolerant programs, our encodings 
are useful to understand how these our two constructs for runtime adaptation 
(a)~may integrate 
into communicating processes 
%that represent an actual programming mechanism.
and (b)~relate to actual programming mechanisms.

\begin{figure}[t!]
\begin{center}
  \begin{tikzpicture}
\node [rectangle,draw] (S1){$S_1$}
       child {node [rectangle,draw] (S2) {$S_2$}
                    child {node [circle,draw] (W3) {$W_3$} edge from parent node[above] {$c$~~~}}
                    child {node [circle,draw] (W4) {$W_4$} edge from parent node[above] {~~~$d$}}
                    edge from parent node[above] {$a$~~~}
                    }
       child {node [circle,draw] (W2) {$W_2$}
       edge from parent node[above] {~~~$b$}};
\end{tikzpicture}
 \end{center}
\caption{A supervision tree: supervisors are depicted as rectangles; workers are represented by circles.}\label{fig:tree}
\end{figure}
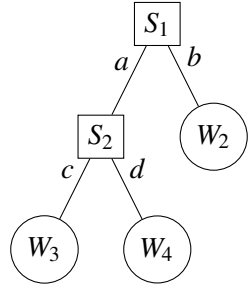

As mentioned in the Introduction, supervision trees organize a program's behavior into two classes: \emph{workers} and \emph{supervisors}. 
Workers are the processes that actually perform computation ---in a com\-mu\-ni\-ca\-tion-centric setting, workers may well correspond to servers which expect and fulfill clients' requests. Supervisors are meant to monitor the behavior of workers by relying on a hierarchical, tree-like structure.
This way, a supervisor may monitor the behavior of a number of children; the supervisor may  itself be supervised by another supervisor at a higher level in the hierarchy. 
As an example, Fig.~\ref{fig:tree} shows a supervisor 
$S_2$ that monitors workers $W_3$ and $W_4$; in turn, $S_2$ is supervised by $S_1$ who is also a supervisor for worker $W_2$.
In Erlang, a supervisor can implement several \emph{restart strategies} which, given a termination signal (e.g. a failure), offer different possibilities for restarting the supervised worker(s). Such strategies are fully detailed in~\cite{erlang:sup}; in what follows, we concentrate on the following two:
%Two of them are the following: %Here we consider only two kinds:
\begin{enumerate}[1.]
 \item \emph{One for one} strategy: if one of the children fails and terminates then only that child is restarted.
 \item \emph{One for all} strategy: if one of the children fails and terminates then all the siblings must be terminated and restarted.
\end{enumerate}

%\jp{We may mention that supervision trees are interesting for us for they are a real (existing) mechanism that is in line for our intended model of adaptation---something that is external to some behavior and may act autonomously for deciding adaptation routines.}

It is interesting to observe how the hierarchical model for fault-tolerance realized by supervision trees in Erlang is in line with our intended model for runtime adaptation:
in addition to 
the loose analogy between located processes as workers and update processes as supervisors, 
we find useful similarities in terms of hierarchical organization of behavior, 
and in the fact that 
both supervision trees and our update processes 
are in essence  \emph{external} recovery mechanisms for precisely delimited behaviors. 
These intuitions are our starting point for the process representations of supervision trees which we analyze in the rest of this section.
%In the following we provide an encoding of supervision trees in our setting, 
For simplicity, we restrict to representing the hierarchy 
given in Fig.~\ref{fig:tree}, assuming that both $S_1$ and $S_2$ realize the same restarting strategy; the representation, however, can be easily generalized to more complex settings. 
We  represent the tree structure by the nesting of located processes, as follows: %, hence obtaining the process:
\begin{equation}\label{eq:trees}
 \componentbig{l_1}{S_1 \para \componentbbig{l_2}{S_2 \para \compo{w_3}{ }{W_3}\compo{w_4}{ }{W_4} } \para \compo{w_2}{ }{W_2}  }
\end{equation}
In our models, we distinguish between localities for supervisors (denoted $l_i$) and localities for workers (denoted $w_j$).
In the following, we describe two different implementations of supervisors, each realizing a different restarting strategy.
%  depending on the analyzed strategy together with an example of a simple worker.

\subsection{Modeling One for one Supervisors}
As suggested by the above informal description, 
the one for one strategy accounts for a form of \emph{local supervision}: 
whenever a child enters in an error state or fails, the child activity must be terminated and the server restarted. 
This way, for instance, in the scenario of Fig.~\ref{fig:tree}, 
if worker $W_3$ fails then its siblings $W_4$ and $W_2$ should be unaffected by (and unaware of) the restarting process that $S_2$ is expected to initiate.

The one for one strategy can be encoded in a quite natural way within our framework. 
Indeed, 
by assuming that workers correspond to servers which are connected to clients, we may represent 
failures as the reception of an invalid value from the client; the restarting process would then correspond to an update action that simply  recreates the located process containing the server.  
More precisely,  we  assume worker $W_3$ 
is an ``arithmetic server'' which
calculates the division between two natural numbers which are communicated by a client $C$ after establishing a session. 
In this context, failure corresponds to division by zero. 
Then, $W_3$ notifies $S_2$ whether it has found a failure (i.e., the denominator being equal to zero). Subsequently,
$S_2$ should decide what to do next: 
(i) if the denominator sent by the client was zero, then there is indeed a failure and  session between $W_3$ and $C$ must terminate immediately;  (ii) otherwise, if there is not a failure, then $S_2$ should allow $W_3$ to perform the division and communicate the result to $C$. 
Observe how in 
 this ``local'' scenario, supervisor $S_2$ does not need to interact with its supervisor $S_1$.
 
To implement the supervision model described above, every time a worker (such as $W_3$) establishes a session with a client (such as $C$), it also connects to its supervisor. This way,  supervisor and worker have a unique communication mechanism.
%has way to communicate its decisions to its worker.
In our case, this mechanism is realized by a session request from $W_3$ to $S_2$ on name $b$. 
By extending the process in (\ref{eq:trees}) with a client  $C$, we obtain 
the following system:
\begin{equation}\label{eq:tree}
 \componentbig{l_1}{S_1 \para \componentbbig{l_2}{S_2 \para \compo{w_3}{ }{W_3}\compo{w_4}{ }{W_4} } \para \compo{w_2}{ }{W_2}  } \para C
\end{equation}
%$$\compo{l_1}{ }{S_1 \para \compo{l_2}{ }{S_2 \para \compo{w_3}{ }{W_3}\compo{w_4}{ }{W_4} } \para \compo{w_2}{ }{W_2}  } \para C$$
where $S_2$,  $W_3$, and  $C$ are defined as follows (the other processes are left unspecified):
$$
\begin{array}{lcl}
S_2&\triangleq& \repopen{b}{\xb}.\inC{\xb}{v}.\select{\xb}{v}.\ifte{v=\fail}{\close{\xb}.\adaptn{w_3}{\compo{w_3}{}{W_3}}}{\close{\xb}} \\ \\
W_3&\triangleq&\repopen{a}{\xa}.\nopenr{b}{\xb}.\inC{\xa}{u}.\inC{\xa}{v}.\ifte{v=0}{\outC{\xb}{\fail}.P}{\outC{\xb}{\ok}.P} \\
P&\triangleq&\branch{\xb}{\fail: \select{\xa}{\fail}.\close{\xa}.\close{\xb} \parallel \ok: \select{\xa}{\ok}.\outC{\xa}{u \div v}.\close{\xa}.\close{\xb}}\\ \\
C&\triangleq&\nopenr{a}{\xa}.\outC{\xa}{c_1}.\outC{\xa}{c_2}.\branch{\xa}{\fail: \close{\xa} \parallel \ok: \inC{\xa}{v}.\close{\xa}}
\end{array}
$$
%\jp{In $S_2$, is it possible to have $\select{\xb}{v}.\close{\xb}.$ inside the if branches, i.e. 
%$$
%\repopen{b}{\xb}.\inC{\xb}{v}.\ifte{v=\fail}{\select{\xb}{v}.\close{\xb} \para \adaptn{w_3}{\compo{w_3}{0}{W_3}}}{\select{\xb}{v}.\close{\xb}}
%$$
%i think it would be a bit more reasonable (and equivalent)}
%the other actors are left unspecified, but in principle their implementation is similar to the processes given above.
%\jp{please explain how the first action of a worker is to "check in" with his supervisor, via a session request}

Observe how process $S_2$ defines an update action on $w_3$ if $W_3$ communicates to him that the invalid value zero was received.
%Importantly, typing ensures that any update on locality $w_3$ can only happen when all the sessions established by $W_3$ are terminated.
Moreover, using our update construct we can implement more interesting supervisors which go beyond service restarting.
For instance, consider the following process: 
$$S'_2\triangleq \repopen{b}{\xb}.\inC{\xb}{v}.\select{\xb}{v}.\close{\xb}.\ifte{v=\fail}{\adaptn{w_3}{\compo{w_3}{0}{W_3'}}}{\nil}$$
Note how $S'_2$ improves over $S_2$ by placing inside $w_3$ a \emph{different} server behavior, represented by $W'_3$.
Suppose that $W'_3$ represents an upgraded version of $W'_3$ in which the arithmetic server no longer relies on a supervisor in
order to deal with failures/invalid values. This way, whenever $W_3$ communicates a failure to $S'_2$, the resulting update action will
lead to an improved server $W'_3$ that will be able to handle the failure locally, without appealing to an external entity.

%where the final update on locality $w_3$ does not just recreate locality $w_3$ with its worker inside but it could provide a different (upgraded) behavior for it. For instance, this feature can be used to encode the following protocol: if the current version of the worker fails (i.e.  a value 0 has been received from the client) then a new  version of the worker is installed. In the new (upgraded) behavior,  upon reception of a value 0, the worker does not communicate the failure to its supervisor  but  sends a special value to the client and complete in a normal state its session.
%Finally, notice that as locality are transparent they could also be updated from an external entity that wishes to change the behavior of both supervisors and workers.\jp{this last sentence is not very clear}

%\jp{A small disadvantage of the example ``the division server gets 0 from the client'' is that it sounds very similar to the naive example for try/catch/exceptions. we should elaborate on this, saying that our example is just for simplicity and that other kinds of behaviors, including those defined externally to the program's text/behavior (say, the deploy two copies of $W_3$) can be handled in our setting.}

Finally, we briefly discuss the typing of previous processes. Given session types
$$
%\begin{array}{lcl}
 \alpha \triangleq  \tin{\integer}.\tin{\integer}.\oplus\{\fail: \epsilon, \ok: \tout{\integer}.\epsilon \}\quad \text{and} \quad 
  \beta \triangleq  \tin{\str}.\oplus\{\fail: \epsilon, \ok:\epsilon\}
%\end{array}
$$
the processes above are well-typed, given the following environments:
$$
\begin{array}{lcl}
 \Gamma&\supseteq& \{ a:\langle \alpha_{\quau}, \overline{\alpha}_{\qual} \rangle, b:\langle \beta_{\quau}, \overline{\beta}_{\qual}\rangle\}\\
 \Theta&\supseteq& \{l_2:\inter{b}{\beta}{\infty} , w_3:\inter{a}{\alpha}{\infty}, \inter{b}{\overline{\beta}}{\infty}\}
\end{array}
$$
In fact, processes $S_2, W_3$, and $C$ can be typed as follows:
$$
\begin{array}{l}
\judgebis{\env{\Gamma}{\Theta}}{S_2}{\type{\emptyset}{\inter{b}{\beta}{\infty}}}\\
\judgebis{\env{\Gamma}{\Theta}}{W_3}{\type{\emptyset}{\inter{a}{\alpha}{\infty}},\inter{b}{\overline{\beta}}{\infty}}\\
\judgebis{\env{\Gamma}{\Theta}}{C}{\type{\emptyset}{\inter{a}{\overline{\alpha}}{1}}}
\end{array}
$$

Notice that as a consequence of being well-typed, the supervisor has to wait that all sessions have been closed before proceeding with an update of its worker.

\subsection{Modeling One for all Supervisors}
While the one for one restarting strategy implements a local relation between a worker and his supervisor, 
in the one for all 
%The second strategy is the symmetric with respect to the previous one, in fact here 
strategy  supervisors need to globally control their children. Hence, whenever one of the children enters in an error state, the child and its siblings have to be terminated and restarted. For instance, in the scenario of Fig.~\ref{fig:tree}, 
if $W_2$ fails then both $W_2$ and all the workers contained in the subtree rooted in $S_2$ need to be terminated and restarted. 
This event should be communicated by $S_1$ to $S_2$, which should handle the termination of its children.
% \jp{I would say: ``It is plausible to think that in such a case $S_1$ would send a signal to $S_2$ bla bla''}. 
On the other hand, if $W_3$ fails then its supervisor $S_2$ only has to stop $W_4$, and nothing needs to be communicated to $S_1$.

The required global control that supervisors should exercise over children in the one for all strategy turns out to be
slightly unnatural to express in our framework of binary communication.
%It is clear, then, that supervisors in a one-for-all strategy need to have a global control on children, and this is somehow less natural to express in our formalism. \jp{I'd add: 
Nevertheless, it is possible to devise a protocol that achieves the desired synchronization sequences.
An alternative to represent global synchronization among services and supervisors is described by the following informal protocol,
in which communication between workers and clients (and between supervisors and workers) is managed via session establishment: 
\begin{enumerate}[1.]
 \item %Each service located in a worker performs one step, i.e. 
 Each worker receives a data value $v$ from a client. The value is checked for validity (via a boolean expression $e(v)$). The result of this check is sent to the worker's supervisor.
 \item Each supervisor waits to receive the result of the validity check from  all its children.
 \item Each supervisor (apart from the root node $S_1$) sends a $\true$ value to its own supervisor. This is to have a uniform communication pattern, and is related to the fact that supervisors themselves cannot fail.
 \item Starting from the root of the tree, a communication cascade takes place: each supervisor communicates to its children if they can continue with their own service or if they have to terminate. Intuitively,  a single worker can continue if and only if (i) none of its siblings have failed and (ii) the supervisor has not received from its own supervisor a failure signal.
 \item If a worker (server) receives a failure signal, then it stops the communications with its clients; otherwise, it can continue with its intended behavior.
 \item Finally, when a worker/server stops it is recreated by its supervisor.  
In the case of a forced termination this corresponds to the actual restarting strategy; in the case of natural termination, this enables server persistency.
In the process $Q$ below, this is realized via an update. Notice that $S_1$ will need to recreate only $W_2$, while $S_2$ will be responsible for the restart of its own children $W_3$ and $W_4$. 
\end{enumerate}

%\jp{similar to a previous comment, this is a bit too focused on failures related with specific data values, so it conveys the idea of exceptions. we should say that the values communicated may not necessarily be data from the client, but system parameters, flags and other infrastructure-related stuff, for instance.}

%\jp{I assume that from now on the text needs to be expanded... to explain the model, for instance}.
Notice that Fig.~\ref{fig:tree} already shows in the edges the names of the services established between each pair. As before,
 we consider a process structure as in (\ref{eq:tree}). %, extended with an instance of a client $C$. %: \jp{the following is redundant, right? it's just like the one before..}
%$$\compo{l_1}{0}{S_1 \para \compo{l_2}{0}{S_2 \para \compo{w_3}{0}{W_3}\compo{w_4}{0}{W_4} } \para \compo{w_2}{0}{W_2}  } \para C$$
Here we just give the encoding of worker $W_3$---the other workers  have analogous representations. 
Moreover, we suppose that $W_3$ can establish a session $i$ with client $C$. 
Notice that each worker can establish a session with just one client at the time.
We have:
$$
\begin{array}{ll}
W_3::=&\nopena{i}{\xj}.\nopenr{c}{\xc}.\inC{\xj}{v}.\ifte{e(v)}{\outC{\xc}{\ok}.P}{\outC{\xc}{\fail}.P} \\
P::=&\inC{\xc}{y}.\select{\xc}{y}.\branch{\xc}{\fail: \select{\xj}{\fail}.\close{\xj}.\close{\xc} \parallel \ok: \select{\xj}{\ok}.\proto_W.\close{\xj}.\close{\xc}}\\ 
C::=& \nopenr{i}{\xj}.\outC{\xj}{c}.\branch{\xj}{\fail: \close{\xj} \parallel \ok: \proto_C.\close{\xj}} \\ \\
% \end{array}
% $$
% 
% $$
% \begin{array}{ll}
S_2::= &\repopen{c}{\xc}.\nopena{d}{\xd}.\nopenr{a}{\xa}.\inC{\xc}{u}.\inC{\xd}{v}.\outC{\xa}{\true}.\inC{\xa}{z}.\select{\xa}{z}.\\
& \ifte{u \wedge v \wedge (z= \ok)}
       {\outC{\xc}{\ok}.\outC{\xd}{\ok}.R}
       {\outC{\xc}{\fail}.\outC{\xd}{\fail}.R}\\
R::= & \branch{\xc}{\fail: \select{\xc}{\fail}.\branch{\xd}{\fail: \select{\xd}{\fail}.\close{\xc}.\close{\xd}.\close{\xa}.Q } \parallel \\
& \qquad \     \ok:\select{\xc}{\ok}.\branch{\xd}{\ok: \select{\xd}{\ok}.\proto_{S2}.\close{\xc}.\close{\xd}.\close{\xa}.Q }}\\
Q::=& \adaptn{w_3}{\compo{w_3}{ }{W_3}} \para \adaptn{w_4}{\compo{w_4}{ }{W_4}} \\ \\
% \end{array}
% $$
% 
% $$
% \begin{array}{ll}
S_1::=&\repopen{a}{\xa}.\nopena{b}{\xb}.\inC{\xa}{u}.\inC{\xb}{v}.\ifte{u \wedge v }
       {\outC{\xa}{\ok}.\outC{\xb}{\ok}.T}
       {\outC{\xa}{\fail}.\outC{\xb}{\fail}.T}\\
T::= & \branch{\xa}{\fail: \select{\xa}{\fail}.\branch{\xb}{\fail: \select{\xb}{\fail}.\close{\xa}.\close{\xb}.\adaptn{w_2}{\compo{w_2}{ }{W_2}} } \parallel \\
& \qquad \     \ok:\select{\xa}{\ok}.\branch{\xb}{\ok: \select{\xb}{\ok}.\proto_{S1}.\close{\xa}.\close{\xb}.\adaptn{w_2}{\compo{w_2}{ }{W_2}} }}
\end{array}
$$

Above, $\proto_W$, $\proto_C$, and $\proto_{Si}$ denote the rest of the behavior of workers, clients and supervisors; as such, we assume that and $\proto_W$ and $\proto_C$ are the dual of each other.
Observe how  as soon as worker $W_3$ has established a session on $i$ with the client, it contacts its own supervisor through $c$ and implements step (1) of the protocol above, then it waits for the decision of the supervisor. Similarly for its supervisor $S_2$: after it has established a session for each worker ($W_3$ and $W_4$) through services $c$ and $d$ respectively, it contacts $S_1$ via $a$ and implements steps (2) and (3) of the protocol. Finally, the root supervisor $S_1$ will take a decision and communicate to its workers whether they have to terminate or continue. Then it will pass the information to (child) supervisor $S_2$, that in turn will rule the behavior of its children (cf. steps (4) and (5)). As a last step (point (6)) each supervisor will take care of recreating of its own children.

The system above can be shown to be well-typed, analogously as we did for the process representation for the one for one strategy.
%Analogously as before the system described here is well-typed, 
Here again, typing ensures that no session is disrupted while communication with clients is still active.

%\todo{Maybe now we need to play a bit with the examples...}\jp{and extend the explanation?}

\section{Concluding Remarks and Future Work}\label{s:concl}
In this short note, we have 
presented the main elements of our work in~\cite{dGP13,dGP13-long}, where we 
have investigated the integration of constructs for 
runtime adaptation into a session types discipline.
Our starting point was our own work on adaptable processes~\cite{BGPZFMOODS},
a process calculi framework for specifying interacting processes which may be suspended, restarted, updated, or discarded at runtime.
Our work in~\cite{dGP13} appears to be the first attempt in addressing the integration
of dynamic evolution issues in models of communicating systems
in which interaction is described (and disciplined) by session types.

The original contributions of this presentation are examples which illustrate how the constructs
for runtime adaptation fit in a session-typed, communication-centric setting. 
We presented and discussed a process representation of Erlang's supervision trees, an existing mechanism
for designing and programming fault-tolerant applications. 
Similarly as update processes, supervisors are external entities which monitor the behavior of one or more workers.
Supervisors are parametric in the restarting strategies that are spawned when a failure occurs in a worker.
We described representations of supervisors into our session process language, and observed that although 
a model of local supervision (in which a supervisor monitors only one worker) is better suited to the communication primitives in our model,
an alternative model of global supervision (in which a supervisor may monitor more than one worker) can also be encoded.
Interestingly, in both cases our session language with runtime adaptation capabilities is able to represent 
more insightful restarting strategies than those provided by the Erlang language. 

The framework in~\cite{dGP13,dGP13-long}, summarized in this note, paves the way for several avenues of future work:
%Next, we elaborate on some of them. 

\begin{enumerate}[-]
\item We would like to 
investigate \emph{progress (deadlock freedom)}
%establish stronger correctness properties
for adaptable, session-typed processes.
This includes adapting to our setting known session type systems for ensuring progress~\cite{DLYTGC07,DBLP:journals/corr/abs-1010-5566}, but also
understanding whether 
the information added by such systems (e.g., orderings on sessions)
can be integrated into  update actions so as to 
prevent/overcome deadlocked sessions at runtime.

%\item Related to the above, here we have described a very simple way of resolving the interplay of communication and evolvability:
%communication actions 
%have priority over update steps.
%As such, the update of a located process is only executed when the sub-processes it contains do not have open sessions.
%This may be inconvenient when already established sessions need to be aborted (as in a session with infinite or deadlocked behavior).
%We would like to refine our current model  
%so as to capture more naturally this kind of scenarios, in which update steps need to be ``forced''.

\item Our nested, transparent locations %(including the treatment of the information on open sessions)
may be too ``open'' for some applications. One may need to enforce
\emph{secure updates}, so as to protect located processes from unintended evolvability actions ---as in, for instance, 
update actions triggered by \emph{unauthorized} users, or update actions 
based on \emph{permissions} for replacing, extending, or destroying the behavior of a located process.
%which completely discard the current behavior without having appropriate permissions to do so.

\item In this presentation, %Finally, as our intention was to explore evolvability issues in known session type structures, 
we have refrained from enriching the typing system with information related to updates.
We think such an extension would not be difficult, and could offer a way to better control update actions.  
This way, for instance, one could decree (and statically check) that 
update actions occur only in selected parts of a communication protocol.
%and statically check that candidate programs attempt update operations accordingly.

\end{enumerate}

%------------------------------------------------------------------------------

\paragraph{Acknowledgments.}
We are grateful to the 
anonymous reviewers: 
their comments and remarks were most useful 
in improving the presentation of this short note. 
Our interest in exploring how supervision trees in Erlang could be modeled in our framework was triggered by a suggestion from
Philip Wadler during PLACES'13 to whom we are grateful. 
This research was supported 
by the Portuguese Foundation for Science and
Technology (FCT) under grants
SFRH\;/\;BPD\;/\;84067\;/\;2012 and CITI, and by French project ANR BLAN 0307 01 - SYNBIOTIC.

%\begin{enumerate}[-]
%\item progress
%\item we privilege communication over updates; a more flexible interplay between update and sessions
%\end{enumerate}

\bibliographystyle{eptcs}
\bibliography{referen}
\end{document}